\documentclass[letterpaper, 9pt, conference]{ieeeconf}
\usepackage{graphicx} 
\IEEEoverridecommandlockouts                              
\usepackage[utf8]{inputenc}
\usepackage[T1]{fontenc}

\usepackage{mathtools, amssymb, bm}

\title{\LARGE \bf Error Bounds Revisited, and How to Use Bayesian Statistics While Remaining a Frequentist}
\author{Ning Xu$^{1}$, Christopher M. Foster$^{2}$, and Jonathan H. Manton$^{3}$
\thanks{$^{1}$Ning Xu is with the Key Laboratory of Advanced Process Control for Light Industry (Ministry of Education), School of Internet of Things Engineering, Jiangnan University, Wuxi 214122, PR China
        {\tt\small ning\_xu12@126.com}}%
\thanks{$^{2}$Christopher M. Foster is with the department of Electrical and Electronic Engineering, The University of Melbourne, Parkville, Australia
        {\tt\small cffoste@student.unimelb.edu.au}}%
\thanks{$^{3}$Jonathan H. Manton is with the Department of Electrical and Electronic Engineering, The University of Melbourne, Parkville, Australia
        {\tt\small j.manton@ieee.org}}%
}

\begin{document}

\maketitle

\newcommand{\R}{\mathbb{R}}

\newcommand{\E}{\bm{E}}

\newtheorem{theorem}{Theorem}
\newtheorem{fact}[theorem]{Fact}
\newtheorem{example}[theorem]{Example}

\begin{abstract}

Signal processing makes extensive use of point estimators and accompanying error bounds. These work well up until the likelihood function has two or more high peaks. When it is important for an estimator to remain reliable, it becomes necessary to consider alternatives, such as set estimators. An obvious first choice might be confidence intervals or confidence regions, but there can be difficulties in computing and interpreting them (and sometimes they might still be blind to multiple peaks in the likelihood). Bayesians seize on this to argue for replacing confidence regions with credible regions. Yet Bayesian statistics require a prior, which is not always a natural part of the problem formulation. This paper demonstrates how a re-interpretation of the prior as a \textit{weighting function} makes an otherwise Bayesian estimator meaningful in the frequentist context. The weighting function interpretation also serves as a reminder that an estimator should always be designed in the context of its intended application; unlike a prior which ostensibly depends on prior knowledge, a weighting function depends on the intended application. This paper uses the time-of-arrival (TOA) problem to illustrate all these points. It also derives a basic theory of region-based estimators distinct from confidence regions.

\end{abstract}

\section{Introduction}

Overwhelmingly, signal processing uses \textit{point} estimators. Error bounds complement point estimates and facilitate ``soft'' decision making. This combination cannot be faulted for well-posed statistical problems when the likelihood function is essentially unimodal.

A multi-modal likelihood function with sharp peaks of a similar height cannot be adequately summarised by a point estimate with an accompanying error bound. Worse still, a typical algorithm will focus on just a single sharp peak, making the point estimate appear to be very precise, when of course there are multiple possibilities for the true parameter corresponding to each of the tall peaks.

Whether the estimate will be used directly by a human, or subsequently processed by another algorithm, we believe it is crucial for the user to be \textbf{made aware of multiple possibilities}.

Another belief is that the choice of an estimator, and accompanying bounds, must be made in the context of its intended usage. It is generally not possible to declare there is a single estimator that should always be used, as strikingly demonstrated by the James-Stein estimator~\cite{James_1961_Estimation,Manton_1998_James-Stein}. For estimating the mean $\mu \in \R^n$ of the Gaussian $x \sim N(\mu,I)$ when the dimension $n \geq 3$, the maximum-likelihood estimate (MLE) $\hat\mu = x$ is not always optimal! Sometimes the James-Stein estimator is favourable \textbf{depending on the application}, or precisely, on how we assess performance.

This paper will therefore develop a basic theory of ``non-point'' estimation in the context of a specific application. The application is searching for a missing person because there is an obvious way of assessing performance: how fast can the person be found?

Finally, the paper presents a pragmatic viewpoint on Bayesian versus frequentist statistics. A Bayesian approach is optimal when performance is being assessed \textit{on average} across all possible parameter values whereas a frequentist approach assesses \textit{individual} performance. Deriving optimal estimators under a frequentist regime can be complicated. We argue that a Bayesian estimator can be interpreted as an ``approximate'' solution to a frequentist problem. Crucially, \textbf{the Bayesian prior should be interpreted as a weighting function}, with peaks in regions where we care most about the estimator's performance.

The connection between these topics will become apparent as the paper progresses.

\textbf{Summary of Contributions:}
\begin{itemize}
    \item Novel theory of region-based estimators.
    \item Novel interpretation of Bayesian prior as a weighting function.
    \item Numerical demonstrations of underlying principles.
\end{itemize}

\textbf{Related work} is given in context throughout the paper.

\section{Weighting Function Estimators}

A version of the well-known Time of Arrival (TOA) problem is the following~\cite{Mallat_2014_Statistics,Zou_2020_Simple}. A person is located at $x \in \R^2$. There are three towers located at $r_1, r_2, r_3 \in \R^2$. The person triggers a beacon at an agreed time and each of the towers records the time at which they observe the beacon, subject to measurement noise $n_i$. Based on the observations $y_i = \| x - r_i \| + n_i$ the aim is to estimate the person's location $x$. Here, $\|\cdot\|$ is the Euclidean norm.

Like most signal processing problems, three types of point estimators have been considered for the TOA problem: MLE~\cite{Xu_2011_Source,Pauley_2018_Existence}, Bayesian MAP~\cite{Viberg_1994_Bayesian,Chen_2016_ToA}, and ``approximate'' estimators based on simplified cost functions designed to aid convergence or reduce computational complexity~\cite{Cheung_2004_Accurate,Xu_2011_Reduced}. After proposing another method for designing a point estimator, the remainder of the paper will focus on set estimators that can reliably report when there is ambiguity in the person's location.

\subsection{A Bayesian-Like Estimator Based on a Weighting Function} \label{sec:weight}

In the frequentist regime, $x$ is ``deterministic but unknown''. The key implication is that performance is assessed on a case-by-case basis. If mean-square error is relevant then the performance of an estimator $\hat x$ is represented by a graph of $\E[ \| \hat x - x \|^2 ; x]$ versus $x$. This leads to the theory of admissible estimators~\cite{Lehmann_1998_Theory}. An estimator is admissible if there is no other estimator which is better than it for all values of $x$.

Finding admissible estimators is difficult. We propose the following sub-optimal approach. (This viewpoint is novel; the outcome is not novel; the difference in interpretation is crucial though, as explained subsequently.) Introduce a pointwise positive \textbf{weighting function} $w(x)$ for the express purpose of reducing the performance graph of $\E[ \| \hat x - x \|^2 ; x]$ versus $x$ to a single number, the average error
\begin{equation} \label{eq:e}
    e = \int w(x) \, \E[ \| \hat x - x \|^2 ; x] \, dx.
\end{equation}
This ``average'' error is not a statistical average because $x$ is not a random variable. It is simply a weighted average where the relative size of $w(x)$ denotes how much we care about the performance of our estimator if the true parameter value is $x$.

\begin{fact} \label{fact:admissible}
    If $w(x) > 0$ pointwise, and $\int w(x)\,dx < \infty$, then an estimator $\hat x$ minimising $e$ in \eqref{eq:e} is admissible.
\end{fact}

This fact is well-known under the guise that Bayesian estimators are admissible; we will get to this connection shortly. It can be verified intuitively by noting that if one function is pointwise lower than another function then its weighted average must also be lower. (Throughout, regularity conditions are ignored for brevity.)

Minimising $e$ in \eqref{eq:e} is easier than it looks. Write $\hat x$ as a function $\hat x(y)$ of the observation $y$. Then $\E[ \| \hat x - x \|^2 ; x] = \int \| \hat x(y) - x \|^2 \, p(y;x) \, dy$ and thus $e = \iint \| \hat x(y) - x \|^2 \, p(y;x) w(x)\,dy\,dx$. Numerically, this evaluates to the same error $e$ as if we 1) pretend $x$ is a random variable with distribution $w(x)$, and 2) compute $\E[\|\hat x - x\|^2]$. Here, we assumed $\int w(x)\,dx = 1$, which is without loss of generality because Fact~\ref{fact:admissible} requires the integral to be finite. In a Bayesian setting, the conditional mean minimises the mean-square error (MSE) $\E[\|\hat x - x\|^2]$.

\begin{theorem}
The Bayesian conditional mean $\hat x(y) = \E[x \mid y]$ minimises the average error $e$ in \eqref{eq:e} when we place on $x$ the prior $p(x) = w(x) / \int w(x')\,dx'$. 
\end{theorem}

This theorem allows us to enjoy the computational benefits of being a Bayesian while remaining a frequentist! It also cautions us \textbf{not to choose the prior in a Bayesian way}, such as based on prior knowledge, or in a non-informative manner.  If individual performance matters then a weighting function $w(x)$ should be used, with the prior becoming $p(x) = w(x) / \int w(x')\,dx'$.

\begin{example}
    In the TOA problem, a Bayesian may well use a prior $p(x)$ based on where people are likely to go hiking. Using our interpretation, we would choose $w(x)$ based on how difficult the terrain is to search. In open areas where visibility is good, we choose $w(x)$ to be small, because as long as the estimate is vaguely correct, it will be easy to spot the person. In areas of dense bush, we choose $w(x)$ to be large because these are the regions where it is hardest to search for the person and thus we want our estimator to be its most accurate.
\end{example}

 Note how the intended application influenced the design of the estimator. When asked to construct a prior, normally one thinks about prior knowledge and ignores the intended application. We thus believe it beneficial to distinguish between a prior and a weighting function despite their mathematical equivalence.

\subsection{Bayesian-like Confidence Regions} \label{sec:cr}

Normally a point estimate suffices for finding a person: start at that point then gradually broaden the search. If the towers are (approximately) collinear though, there is (approximate) ambiguity: to each point on one side of the towers there is a corresponding point on the other side for which the time-of-arrival measurements would be (approximately) identical. Searching for a missing person based on a point estimator could be disastrous. Instead, two search teams should be dispatched, one for each of the two sides.

There are different ways to convey where to look for a missing person. A finite set can be given. Information is lost though unless it is equally likely for the person to be near any of these points.

We propose using an estimator which returns a set, thought of as search regions, where the plural ``regions'' implies the set need not be connected. Set estimators have a long history, and include confidence regions and credible regions, so by ``propose'' we really mean three things.
\begin{enumerate}
    \item We argue for wider usage of set estimators.
    \item We emphasise the need to consider the intended application.
    \item We give suggestions for implementing such estimators.
\end{enumerate}

Before presenting a basic theory, we explain briefly why we are not content with traditional error bounds or confidence intervals. Error bounds obtained by examining the likelihood function locally, such as the Cramer-Rao Bound, by definition are blind to multiple high peaks of the likelihood function. Confidence intervals, by definition, return a connected interval and can thus be much wider than desired if additional high peaks are present. Their generalisation to disconnected confidence regions can overcome this, but often their derivation does not account for the intended application. This can lead to poor performance~\cite{Morey_2016_fallacy, Welch_1939_Confidence}. Finally, computing disconnected confidence regions can be far from trivial, and confidence regions are not unique. Even finding optimal confidence intervals for multi-dimensional Gaussian distributions is difficult~\cite{Tseng_1997_Good}. In short, it is inadequate to say ``use confidence regions'' and leave it at that.

Consider again the TOA problem in the context of finding a missing person. It is a frequentist problem because we care about finding each and every individual person regardless of where they went missing. In other words, our measure of performance will not involve a prior. If our set estimator is $A(y)$ then a possible performance measure is $\inf_{x \in X} \Pr[x \in A(y); x]$ where $X \subset \R^2$ represents the largest possible region the person might be; choosing $X$ to be compact accords with Earth having finite surface area.

This performance measure is the starting point for confidence regions: for $0 < c \leq 1$, if the sets $A_c(y)$ satisfy $\inf_{x \in X} \Pr[x \in A_c(y); x] \geq c$ then $A_c(y)$ are confidence regions at level $c$. For a fixed $c$, the established theory of confidence regions often deems it desirable to choose each $A_c(y)$ to have the smallest area~\cite{Hyndman_1996_Computing}. But this ignores the intended application!

Assume some regions of $X$ are easier to search than others; some parts are flat and open while other parts are hilly and forested. Then the time needed to search the region $A \subset X$ can be taken to be
\begin{equation} \label{eq:nu}
    \nu(A) = \int_A v(x)\,dx
\end{equation}
where $v$ is a non-negative function whose value $v(x)$ is proportional to the difficulty with which the location $x$ can be searched. (Actually, $\nu(A)$ is a measure and $v(x)$ its Radon-Nikodym derivative with respect to Lebesgue measure $dx$.)

Finding regions $A_c(y)$ satisfying $\inf_{x \in X} \Pr[x \in A_c(y); x] \geq c$ and minimising $\nu(A_c(y))$ is very challenging. We propose applying a similar trick to that used in \S\ref{sec:weight}.
Relax the individual constraints $\Pr[x \in A_c(y); x] \geq c$ to the weighted average constraint
\begin{equation} \label{eq:weight2}
    \int_{x \in X} w(x) \, \Pr[x \in A_c(y); x] \, dx \geq c \int_{x \in X} w(x)\,dx
\end{equation}
where $w(x)$ is a non-negative weighting function. Increasing $w(x)$ at $x$ increases the reliability of the confidence regions $A_c(y)$ if the true location of the person is $x$. Numerically, \eqref{eq:weight2} is equivalent to $\Pr[x \in A_c(y)] \geq c$ where $x$ is being treated as a random variable with distribution $p(x) = w(x) / \int w(x)\,dx$; see the Appendix.

In Bayesian statistics, regions $A_c(y)$ satisfying $\Pr[x \in A_c(y)] \geq c$ are called credible regions. We have thus transformed a frequentist problem into a numerically Bayesian problem while remaining a frequentist. The distinction is we understand from \eqref{eq:weight2} that the ``prior'' is actually a weighting function controlling where the confidence regions are going to be more reliable.

\subsection{Credible Regions}

There are many regions $A_c(y)$ satisfying $\Pr[x \in A_c(y)] \geq c$. One common choice is minimising the areas of the $A_c(y)$, a choice known as the Highest Density Region (HDR). Such regions have the form $A_c(y) = \{ x \mid p(x | y) \geq \alpha_y \}$ where $\alpha_y \in \R$ is the largest possible value for which $\Pr[x \in A_c(y)] \geq c$ holds.

Our two criticisms of directly using credible regions are
\begin{enumerate}
    \item the intended application is not taken into account;
    \item there is no guidance for how to choose the prior in a frequentist setting.
\end{enumerate}
These criticisms were addressed in the previous section. Rather than measure area using the Lebesgue measure we should use an application-specific measure \eqref{eq:nu}. And we should replace the prior by a weighting function whose interpretation is clear from \eqref{eq:weight2}.

All that remains is to incorporate the application-specific measure $\nu(A)$ into the theory of credible regions.

\begin{theorem} \label{th:cred}
    Define the measure $\nu(\cdot)$ as in \eqref{eq:nu}. Let
    \begin{equation}
        A_c(y) = \Big\{ x \mathrel{\Big|} p(x \mid y) \, \frac1{v(x)} \geq \alpha_c \Big\}
    \end{equation}
    where $\alpha_c \in \R$ is the largest value for which
    \begin{equation} \label{eq:pxyc}
        \int_{A_c(y)} p(x \mid y) \, dx \geq c.
    \end{equation}
    Provided $\alpha_c > 0$ then $A_c(y)$ has the smallest volume $\nu(A_c(y))$ out of any region satisfying \eqref{eq:pxyc}.
\end{theorem}

The above theorem shows that the credible regions are given by the level sets of $p(x \mid y) / v(x)$. See the Appendix for a proof.

\subsection{Growth of Credible Regions}

The credible regions $A_c(y)$ defined in the previous section grow monotonically with $c$. Reporting the regions $A_c(y)$ for different values of $c$ provides richer information to the search teams. Conceptually, the search teams should start by searching the region $A_c(y)$ for $c$ small, then if the person is not found, increase $c$ and search the not previously searched portions of the updated $A_c(y)$.

One way to visualise how $A_c(y)$ changes with $c$ is to draw its boundary for different values of $c$. From Theorem~\ref{th:cred} this is equivalent to drawing the contours of $p(x \mid y) / v(x)$.

\section{The Weighting Function}

For both point estimators and set estimators, we have proposed an approach for constructing an estimator that depends on a weighting function. Changing the weighting function changes the estimator.

The role of the weighting function can be understood in the context of multi-objective optimisation. We have an infinite number of objectives: minimise the error for each value of the true parameter. As we try to optimise one objective we are likely to make another worse. Changing the weighting function allows us to change the trade-offs between these objectives. (Precisely, we can sweep over the Pareto front.)

If necessary, the weighting function can be tuned with the aid of simulations. Start with a relatively flat weighting function, simulate the performance of the estimator, then increase the value of the weighting function in regions where the estimator performs unacceptably poorly, and repeat.

\section{Numerical Demonstration}

In this simulation, the task is to search for a missing person within a $20$ by $20$ unit area. This search area includes three towers located at $(5,2)$, $(1,10)$, and $(15,7)$, which operate with a measurement noise following $\mathcal{N}(0,\sigma^2$). The missing person is assumed to be walking along a path defined by the line $y=x$. There are two high-risk regions that are difficult to search: a bush area centred at $(5,5)$ with a radius of $2\sqrt{2}$, and a forest area centred at $(12,12)$ with a radius of $3\sqrt{2}$. 

\begin{figure}[!hbt]
\centering
\includegraphics[width=\columnwidth]{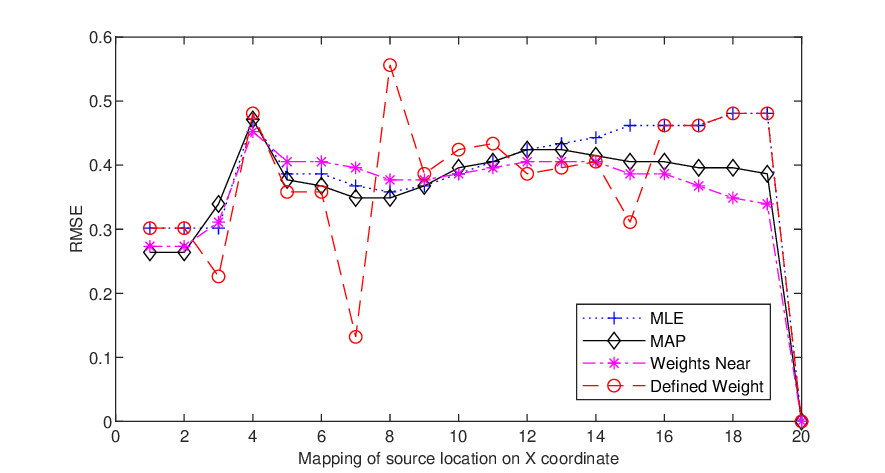}
\caption{The RMSE of various estimators versus the source position}
\label{fig1}
\end{figure}

Fig.~\ref{fig1} compares the root mean square error (RMSE) for different estimators at various source positions along the described path. The MLE estimator is equivalent to using a uniform weighting function that provides no information on the environment. The MAP estimator employs a Bayesian prior, which is defined as the sum of two Gaussian distributions based on the fact that two high-risk areas are more likely to have hikers. The "Weights Near" estimator uses a weighting function based on the distance to the path $y=x$, representing the scenario where the prior information about the person’s known walking route is incorporated. The "Defined Weights" estimator utilises an environment-based weighting function, which is inversely proportional to the distance from the centres of the two high-risk areas, indicating the difficulty of searching the terrain. This last method is meant to operate similar to a Bayesian conditional mean as described in Theorem 2.  Fig. \ref{fig1} shows that both methods derived from constructing weighting functions based on intended application produce admissible estimators. 

\begin{figure}[!hbt]
\centering
\includegraphics[width=\columnwidth]{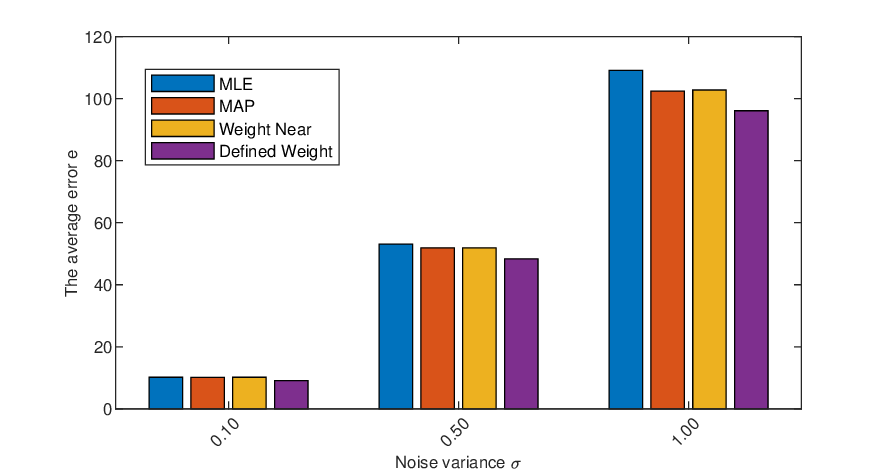}
\caption{The average error $e$ of various estimators under different $\sigma$}
\label{fig2}
\end{figure}

 What is of even more particular interest is the fact that performance seems to improve for both the "Weights Near" and "Defined Weights" methods within the high-risk regions. The fact that these estimators operate poorly in the regions outside of the high-risk regions is intended, as both are operating under the assumption that it will be easier to search those environments (such as an open plain or field). Using \eqref{eq:e}, the average error $e$ for the estimators above was computed and shown in Fig.~\ref{fig2}. This figure shows the average error for each estimator under differing noise variances. These results indicate that the defined weighting function achieves superior desired performance when compared to other approaches in terms of minimising the defined error $e$.

\begin{figure}[!hbt]
\centering
\includegraphics[width=\columnwidth]{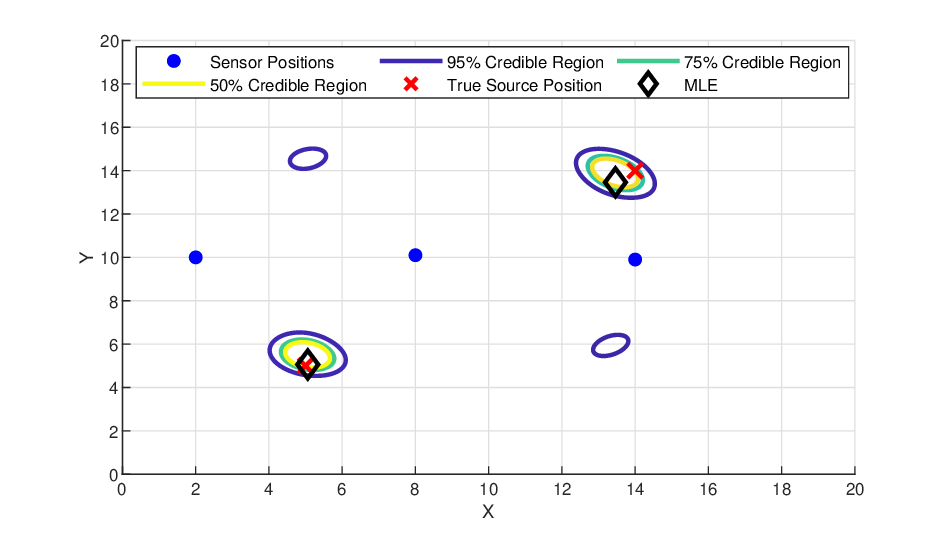}
\caption{The credible region $A_c(y)$ under different $c$}
\label{fig3}
\end{figure}

Fig.~\ref{fig3} illustrates the variation in the credible region of the Bayesian-like estimator employing the defined weighting function as the confidence level $c$ changes. This scenario assumes that the towers are positioned approximately  collinear to each other at locations $(2,10)$, 
 $(8,10.1)$, and $(14,9.9)$. The results indicate that when individuals are potentially lost in high-risk regions, the defined weighting function prioritises those regions. This results in a notable change in the credible regions. Whereas a point estimator would be unable to truly distinguish between either side of the tower line, employing the weighting function allows for more emphasis to be placed in the high-risk regions. This alters the credible region such that the other side of the towers is not even considered until the confidence level is increased significantly. As the confidence level $c$ increases, the credible regions $A_c(y)$ expands, transitioning from a single region into two disjoint regions. This behaviour reflects the scenario where a lost person is not found quickly and the search area must be expanded. The transition to include a second region reflects an acknowledgement that the signal could have originated from the other side of the tower line and that the area should be searched as well to ensure that the person is found.

\section{Conclusion}

Bayesian statistics has advantages over frequentist statistics but, in certain applications, the prior can be troubling for a frequentist and incorrectly chosen by a Bayesian. Re-interpreting the prior as a weighting function (\S\ref{sec:weight}) offers a middle ground and serves as a reminder to always consider the intended application when designing an estimator. This idea extends to set estimators, where frequentist approaches can be difficult to derive or difficult to interpret. Building on this, novel guidelines for deriving and implementing set estimators were derived in \S\ref{sec:cr}. Using set estimators more broadly in signal processing is strongly advocated, especially when the likelihood function might be multi-modal with tall narrow peaks of relatively similar heights. Failing to do so can result in the potentially dangerous situation where an estimator is thought to be accurate but is grossly inaccurate.




\section*{Appendix}

\textbf{Manipulation of \eqref{eq:weight2}:}
Let $p(x) = w(x) / \int w(x) \, dx$. Then \eqref{eq:weight2} becomes $\int p(x)\,\Pr[x \in A_c(y);x]\,dx \geq c$. The left-hand side (LHS) equals $\int p(x) \int 1_{\{x \in A_c(y)\}} \, p(y;x) \, dy \, dx$ where $1_{\{\cdot\}}$ is the indicator function. Promoting $x$ to a random variable means $p(y;x) p(x)$ becomes $p(y \mid x) p(x) = p(x,y)$. The LHS is thus $\Pr[x \in A_c(y)]$.

\textbf{Proof of Theorem~\ref{th:cred}:}
Fix a $y$ and let $q(x) = p(x \mid y)$. Note that $\int_A q(x)\,dx = \int_A q(x)/v(x)\,v(x)\,dx = \int_A q(x) / v(x) \, d\nu$. Fix an $\alpha > 0$ and let $A = \{ x \mid q(x) / v(x) \geq \alpha\}$ and $c = \int_A q(x)/v(x)\,d\nu$. It suffices to show that if $B$ is any set for which $\int_B q(x)/v(x)\,d\nu \geq c$ then $\nu(B) \geq \nu(A)$. Write $B = (B \cap A^\mathsf{c}) \cup (B \cap A)$ where $A^\mathsf{c}$ is the set complement of $A$. On $A$ we have $q(x) / v(x) \geq \alpha$, thus $\int_{A \cap B^\mathsf{c}} q(x)/v(x)\,d\nu \geq \alpha \, \nu(A \cap B^\mathsf{c})$. On $B \cap A^\mathsf{c}$ we have $q(x)/v(x) < \alpha$, thus $\int_{B \cap A^\mathsf{c}} q(x)/v(x)\,d\nu \leq \alpha\, \nu(B \cap A^\mathsf{c})$. Now,
\begin{align}
    c &\leq \int_B q/v \,d\nu \\
    &= \int_A q/v \, d\nu - \int_{A \cap B^\mathsf{c}} q/v \, d\nu + \int_{B \cap A^\mathsf{c}} q/v \,d\nu \\
    &\leq c - \alpha\,\nu(A \cap B^\mathsf{c}) + \alpha\,\nu(B \cap A^\mathsf{c}).
\end{align}
Since $\alpha > 0$, we have $\nu(A \cap B^\mathsf{c}) \leq \nu(B \cap A^\mathsf{c})$ and hence
\begin{align}
    \nu(B) &= \nu(B \cap A^\mathsf{c}) + \nu(B \cap A) \\
        &\geq \nu(A \cap B^\mathsf{c}) + \nu(B \cap A) \\
        &= \nu(A),
\end{align}
as required. (The case $\alpha=0$ is uninteresting because it corresponds to $c=1$.)


\bibliographystyle{abbrv}
\bibliography{reference.bib}

\begin{thebibliography}{10}

\bibitem{Chen_2016_ToA}
X.~Chen, S.~Song, and J.~Xing.
\newblock A {{ToA}}/{{IMU}} indoor positioning system by extended {{Kalman}}
  filter, particle filter and {{MAP}} algorithms.
\newblock In {\em 2016 {{IEEE}} 27th {{Annual International Symposium}} on
  {{Personal}}, {{Indoor}}, and {{Mobile Radio Communications}} ({{PIMRC}})},
  pages 1--7, Sept. 2016.

\bibitem{Cheung_2004_Accurate}
K.~Cheung, W.~Ma, and H.~So.
\newblock Accurate approximation algorithm for {{TOA-based}} maximum likelihood
  mobile location using semidefinite programming.
\newblock In {\em 2004 {{IEEE International Conference}} on {{Acoustics}},
  {{Speech}}, and {{Signal Processing}}}, volume~2, pages ii--145, May 2004.

\bibitem{Hyndman_1996_Computing}
R.~J. Hyndman.
\newblock Computing and {{Graphing Highest Density Regions}}.
\newblock {\em The American Statistician}, 50(2):120--126, May 1996.

\bibitem{James_1961_Estimation}
W.~James and C.~Stein.
\newblock Estimation with quadratic loss.
\newblock In {\em In {{Proc}}.4th {{Berkeley Symp}}. {{Math}}. {{Statist}}.},
  volume~1, pages 311--319. in Proc.4th Berkeley Symp. Math. Statist., 1961.

\bibitem{Lehmann_1998_Theory}
E.~L. Lehmann and G.~Casella.
\newblock {\em Theory of Point Estimation}.
\newblock Springer Texts in Statistics. Springer, New York, 2nd ed edition,
  1998.

\bibitem{Mallat_2014_Statistics}
A.~Mallat, S.~Gezici, D.~Dardari, C.~Craeye, and L.~Vandendorpe.
\newblock Statistics of the {{MLE}} and {{Approximate Upper}} and {{Lower
  Bounds-Part I}}: {{Application}} to {{TOA Estimation}}.
\newblock {\em IEEE Trans. Signal Process.}, 62(21):5663--5676, Nov. 2014.

\bibitem{Manton_1998_James-Stein}
J.~Manton, V.~Krishnamurthy, and H.~Poor.
\newblock James-{{Stein}} state filtering algorithms.
\newblock {\em IEEE Trans. Signal Process.}, 46(9):2431--2447, Sept. 1998.

\bibitem{Morey_2016_fallacy}
R.~D. Morey, R.~Hoekstra, J.~N. Rouder, M.~D. Lee, and E.-J. Wagenmakers.
\newblock The fallacy of placing confidence in confidence intervals.
\newblock {\em Psychon Bull Rev}, 23(1):103--123, Feb. 2016.

\bibitem{Pauley_2018_Existence}
M.~Pauley and J.~H. Manton.
\newblock The {{Existence Question}} for {{Maximum-Likelihood Estimators}} in
  {{Time-of-Arrival-Based Localization}}.
\newblock {\em IEEE Signal Processing Letters}, 25(9):1354--1358, Sept. 2018.

\bibitem{Tseng_1997_Good}
Y.-L. Tseng and L.~D. Brown.
\newblock Good exact confidence sets for a multivariate normal mean.
\newblock {\em Ann. Statist.}, 25(5), Oct. 1997.

\bibitem{Viberg_1994_Bayesian}
M.~Viberg and A.~Swindlehurst.
\newblock A {{Bayesian}} approach to auto-calibration for parametric array
  signal processing.
\newblock {\em IEEE Trans. Signal Process.}, 42(12):3495--3507, Dec. 1994.

\bibitem{Welch_1939_Confidence}
B.~L. Welch.
\newblock On {{Confidence Limits}} and {{Sufficiency}}, with {{Particular
  Reference}} to {{Parameters}} of {{Location}}.
\newblock {\em Ann. Math. Statist.}, 10(1):58--69, Mar. 1939.

\bibitem{Xu_2011_Reduced}
E.~Xu, Z.~Ding, and S.~Dasgupta.
\newblock Reduced {{Complexity Semidefinite Relaxation Algorithms}} for
  {{Source Localization Based}} on {{Time Difference}} of {{Arrival}}.
\newblock {\em IEEE Trans. on Mobile Comput.}, 10(9):1276--1282, Sept. 2011.

\bibitem{Xu_2011_Source}
E.~Xu, Z.~Ding, and S.~Dasgupta.
\newblock Source {{Localization}} in {{Wireless Sensor Networks From Signal
  Time-of-Arrival Measurements}}.
\newblock {\em IEEE Trans. Signal Process.}, 59(6):2887--2897, June 2011.

\bibitem{Zou_2020_Simple}
Y.~Zou and H.~Liu.
\newblock A {{Simple}} and {{Efficient Iterative Method}} for {{Toa
  Localization}}.
\newblock In {\em {{ICASSP}} 2020 - 2020 {{IEEE International Conference}} on
  {{Acoustics}}, {{Speech}} and {{Signal Processing}} ({{ICASSP}})}, pages
  4881--4884, May 2020.

\end{thebibliography}

\end{document}